\documentclass[runningheads]{llncs}
\usepackage{graphicx}
\usepackage{tabularx}
\begin{document}
\title{Future Unruptured Intracranial Aneurysm Growth Prediction using Mesh Convolutional Neural Networks }
\titlerunning{Future UIA Growth Prediction with MeshCNNs}
%
\author{Kimberley M. Timmins\inst{1} 
\and
Maarten J. Kamphuis\inst{2} 
\and
Iris N. Vos\inst{1} 
\and
Birgitta  K. Velthuis\inst{2}
\and
Irene C. van der Schaaf* \inst{2}  
\and
Hugo J. Kuijf* \inst{1}} 
\authorrunning{K.M.Timmins et al.}
%
\institute{{Image Sciences Institute, University Medical Center Utrecht, Utrecht, The Netherlands.}
\email{k.m.timmins@umcutrecht.nl}\and
{Department of Radiology, University Medical Center Utrecht, Utrecht, The Netherlands.} \\ 
\texttt{*Joint last author}
}

\maketitle              
\begin{abstract}

The growth of unruptured intracranial aneurysms (UIAs) is a predictor of rupture. Therefore, for further imaging surveillance and treatment planning, it is important to be able to predict if an UIA is likely to grow based on an initial baseline Time-of-Flight MRA (TOF-MRA).  It is known that the size and shape of UIAs are predictors of aneurysm growth and/or rupture.  We perform a feasibility study of using a mesh convolutional neural network for future UIA growth prediction from baseline TOF-MRAs. We include 151 TOF-MRAs, with 169 UIAs where 49 UIAs were classified as growing and 120 as stable, based on the clinical definition of growth (\textgreater 1 mm increase in size in follow-up scan). UIAs were segmented from TOF-MRAs and meshes were automatically generated. We investigate the input of both UIA mesh only and region-of-interest (ROI) meshes including UIA and surrounding parent vessels. We develop a classification model to predict UIAs that will grow or remain stable. The model consisted of a mesh convolutional neural network including additional novel input edge features of shape index and curvedness which describe the surface topology.  It was investigated if input edge mid-point co-ordinates influenced the model performance.  The model with highest AUC (63.8\%) for growth prediction was using UIA meshes with input edge mid-point co-ordinate features (average F1 score = 62.3\%, accuracy = 66.9\%, sensitivity = 57.3\%, specificity = 70.8\%). We present a future UIA growth prediction model based on a mesh convolutional neural network with promising results.

\keywords{meshes \and aneurysms \and growth prediction \and geometric deep learning \and topology.}
\end{abstract}
\section{Introduction}

Approximately 3\% of the general population has a unruptured intracranial aneurysm (UIAs) \cite{Greving2014DevelopmentStudies}. If an UIA ruptures, it leads to subarachnoid haemorrhage with a high mortality and morbidity rate. Neurosurgical or endovascular treatment can prevent UIAs from rupture, but carry a considerable risk. Therefore a balanced decision based on the rupture and treatment complication risk must be made \cite{Algra2019ProceduralMeta-analysis}. UIA growth is an important rupture risk factor \cite{vanderKamp2021RiskGrowth}, and if detected, preventative treatment should be considered. Most UIAs are monitored, using Time-of-Flight Magnetic Resonance Angiographs (TOF-MRAs) or Computed Tomography Angiographs (CTAs). Currently, 2D size measurements of the UIAs are made and an aneurysm will be considered to be growing if there is a change in size (\textgreater 1 mm) \cite{Hackenberg2019DefinitionGroup}. UIA shape is also known to be different in aneurysms that grow \cite{Backes2017}. 
The ELAPSS score \cite{Backes2017} is a clinical score for UIA growth prediction based on patient and aneurysm characteristics. The predictors are: Earlier subarachnoid hemorrhage, aneurysm Location, Age, Population, aneurysm Size and Shape. Shape is assessed visually as 'Regular' or 'Irregular'. \

As computer-aided radiology tools continue to be developed, 3D volume and morphology measurements of UIAs can be made \cite{Timmins2021ReliabilityAneurysms}, including to distinguish between growing and stable aneurysms \cite{Leemans2019IntracranialChanges,Leemans2019ComparingAneurysms,Timmins2022RelationshipAneurysms}. UIA rupture risk prediction models have been developed based on morphological parameters, as well as classical parameters \cite{Kim2019MachineAneurysm,Liu2018PredictionNetwork}. More recently, some prediction models for aneurysmal stability and growth have been proposed \cite{Liu2019PredictionFeatures,Bizjak2021DeepGrowth}. \

Liu et al.\cite{Liu2019PredictionFeatures} investigated predicting aneurysm stability using machine learning regression models and 12 morphology radiomics features. The dataset included 420 aneurysms (4 - 8 mm). Instability was defined as ruptured within a month, growth or adjacent structure compressive symptoms. They determined flatness to be the most important morphological predictor of aneurysm stability. Bizjak et al.\cite{Bizjak2021DeepGrowth} found using point clouds with PointNet++ for future UIA growth prediction had a higher accuracy than other machine learning models based on morphological parameters. The method was performed using only 44 UIAs, where 25 growing and 19 stable. UIAs were visually inspected in 3D to be classified as growing or stable. \

Various different morphology measurements and definitions of growth or stability have been used in these studies, making it difficult to make direct comparisons. However, it is clear that UIA shape and surface topology is an important predictor of future UIA growth and that deep learning methods may have an advantage over using predefined morphology parameters. Geometric deep learning methods are well suited to this problem, as they accurately describe the shape and topology of a surface by using point clouds or meshes \cite{Cao2020ALearning}. Meshes may have a preference over point clouds as they include connectivity information, providing more information about the surface topology. Segmented UIA meshes could be used as we already know UIA shape is a growth predictor growth. Alternatively, parent vessels in a Region-of-Interest (ROI) around the UIA could be used which includes UIA-vessel configuration and exact UIA segmentation is not required. \

MeshCNN\cite{Hanocka2019} is a convolutional neural network (CNN) developed for classification and segmentation problems using 3D triangular meshes. Convolutions and pooling are performed on edges of the meshes, based on an edge neighbourhood. Five relative scale, translation and rotation invariant geometric edge are determined for each edge as input features for the model. These five geometric features are: the dihedral angle, two inner angles and two edge-length ratios. MeshCNN has only been used for a few medical imaging classification and segmentation problems, including age prediction based on the neonatal white matter cortical surface\cite{Vosylius2020GeometricSurface} and UIA segmentation from a parent vessel\cite{Schneider2021MedmeshCNNModels}. In our previous work, we proposed a modified version of MeshCNN for UIA detection based on brain vessel surface meshes\cite{Timmins2022DeepDetection}. \\

In this paper, we propose a prediction model for future UIA growth from baseline TOF-MRAs using a mesh convolutional neural network. We investigate the use of meshes of UIAs alone, and region-of-interest (ROI) meshes including the UIA with parent vessels as input for these models and their performance for future UIA growth prediction. We also investigate the addition of edge mid-point co-ordinate input features of the meshes and the impact on the model performance. 

\section{Materials and Methods}
\subsection{Dataset}
The dataset consisted of 151 baseline Time-of-Flight MRAs (TOF-MRAs) taken from routine clinical scans. We included patients with UIAs who met the following inclusion criteria: 1) A TOF-MRA or CTA was available at baseline and follow-up, 2) the follow-up scan was performed at least 6 months after the baseline scan, and 3) the patient had at least 1 untreated UIA present on both baseline and follow-up imaging. The most recent follow-up scan in which the UIA remained untreated and unruptured was used for growth assessment. Fusiform and ruptured aneurysms were excluded. All scans were made from the University Medical Center Utrecht between 2006 and 2020. The average time between baseline and follow-up scans was 5.2 $\pm$  3.3 years (range: 1 - 16 years)   The mean baseline aneurysm size was 5.0 $\pm$ 2.2 mm with a range of 1.3 – 14.7 mm. 
Manual 2D length and width UIA measurements were performed in IntelliSpace Portal (Philips Healthcare) by an experienced neuro-radiologist (I.C.v.d.S.) and a trained PhD-student (M.J.K.) according to standard clinical protocol. Growth was defined as a $\ge$ 1.0 mm increase in any direction between the baseline and follow-up scan\cite{Hackenberg2019DefinitionGroup}. Based on this definition, UIAs were categorised as either ‘growing’ (30\%, n= 49) or ‘stable’ (70\%, n = 120). 

\subsection{Methods}

\subsubsection{Input Mesh Generation}
All baseline TOF-MRAs were pre-processed using an N4 bias field correction algorithm and z-score normalised before being resampled to have voxel size 0.357 mm x 0.357 mm x 0.500 mm (median of the dataset). All UIA and ROI selection, mesh generation and processing was performed completely automatically based on UIA annotations. 

\paragraph{UIA mesh generation}
UIA meshes were generated and pre-processed automatically based on the TOF-MRAs and UIA annotations. UIAs were manually segmented from the TOF-MRAs using annotations drawn on axial slices in in-house-developed software implemented in MeVisLab (MeVis Medical Solutions) (performed by I.C.v.d.S. and M.J.K). A triangular mesh was automatically fitted to the outside of the UIA surface using a Marching Cubes algorithm \cite{Lorensen1987}. All UIA meshes were down-sampled to 1000 edges and included just the UIA and no other vessels.

\paragraph{Region-of-Interest (ROI) mesh generation}
ROI meshes were automatically generated from the TOF-MRAs using the UIA segmentations. An existing 3D U-net was used to automatically perform full vessel segmentation from the scans\cite{deVos2021AutomaticLearning}. Based on the UIA segmentation, a region-of-interest (ROI) including only the UIA and parent vessels was made. The centre-of-mass of the UIA segmentation was determined and the ROI included all connected vessels (and UIA) within a 20 mm cube around the centre-of-mass. A mesh was automatically fitted to the outside of the UIA and parent vessel surface using a Marching Cubes algorithm \cite{Lorensen1987}. All ROI meshes were down-sampled to 2000 edges. 

\paragraph{Input edge features} 
Based on the generated UIA and ROI meshes, new input edge features were automatically determined per edge. These were shape index, curvedness and edge mid-point co-ordinates. These further edge features (shape index, curvedness and mid-point co-ordinates) could then be included as input to the network, in addition to the original five geometric edge features. \

Shape index and curvedness are rotation and translation invariant measures which describe the topology of the UIA surface. The invariant nature of these novel input edge features ideal for use in MeshCNN. It is known from our previous work in UIA detection that the addition of both shape index and curvedness as input edge features improve the performance of the original MeshCNN \cite{Timmins2022GeometricDetection}. Shape descriptor values; shape index and curvedness, were calculated for each vertex on the mesh surface using the standard formulae \cite{Koenderink1992}.  An edge was then given a shape descriptor value (shape index or curvedness), as being the average of the values at the corresponding end vertices of the edge. \

The addition of edge mid-point co-ordinate values was suggested in the original MeshCNN paper \cite{Hanocka2019}. We experiment with including these co-ordinates in our models as we know location is important as an aneurysm growth predictor \cite{Backes2017}. Edge mid-point co-ordinates (x,y,z) were determined as the average of the world co-ordinates of the corresponding end vertices of the edge. \

\begin{figure}
\includegraphics[width=\textwidth]{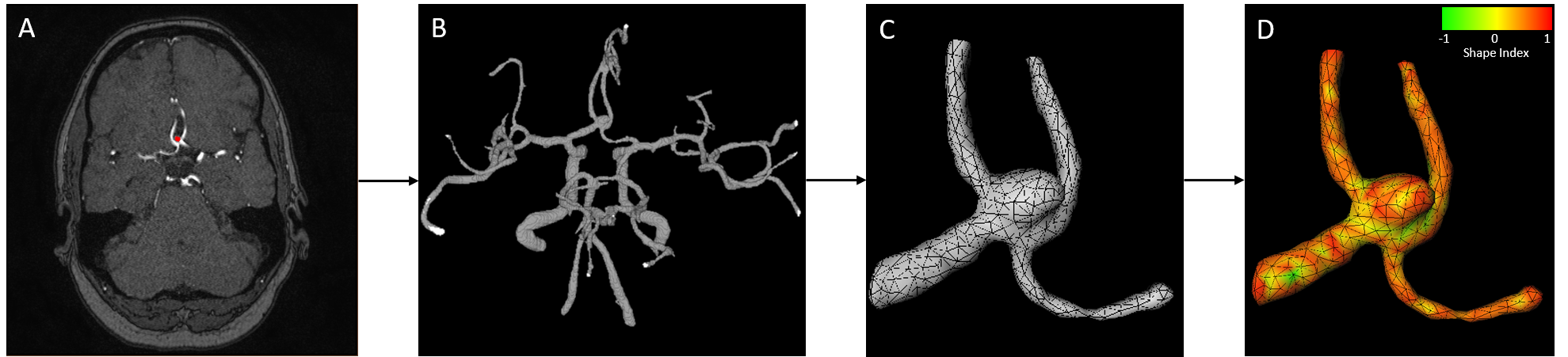}
\caption{Example generation of input region-of-interest (ROI) mesh including parent vessels and UIA. A: TOF-MRA with annotated UIA shown overlaid in red. B: Vessel segmentation performed using 3D U-net \cite{deVos2021AutomaticLearning}. C: ROI selection including UIA and parent vessels, followed by mesh generation. D: Shape index determination for each edge, to be used as an additional input feature alongside curvedness and edge coordinates. } \label{roi_mesh}
\end{figure}

Figure \ref{roi_mesh} shows an example generation of a ROI mesh including shape index values determined for each edge.

\subsubsection{Model Implementation}
A ConvNet style network was set up based on our modified MeshCNN framework \cite{Timmins2022DeepDetection} including four convolutional layers and four pooling layers.  Four different model configurations were investigated. The first model (uia\_model) had UIA meshes only as input, with 1000 edges. Pooling layer configuration for the UIA model was: 750, 600, 500, 400. The second model (roi\_model) had ROI meshes including UIA and parent vessels as input. The pooling layer configuration for the ROI model was: 1500, 1200, 1000, 800. \
All models were made to include shape index and curvedness as additional input features to the original five edge geometric features of MeshCNN. This meant that there were seven input edge features as standard. For each different input, two models were trained. The first with the seven input edge features (uia\_model\_1, roi\_model\_1), and the second including edge mid-point co-ordinates (x,y,z) as further additional input features (uia\_model\_2, roi\_model\_2), meaning there were ten input edge features. No augmentation was used.\

For all models, all other hyperparameters were kept the same, and as similar to the original paper as possible\cite{Hanocka2019}.  Both a weighted data sampler and weighted cross-entropy loss function were used, based on the class distribution of growing and stable UIAs (0.7 to growing, 0.3 to stable). Batch normalisation was used with a batch size of 50 meshes and a learning rate of 0.0002. 
The classification model was trained to predict future growth of the UIA as defined by the clinical definition, whereby output was one of the two classes: growing or stable. All experiments were performed using five-fold cross-validation where the validation splits were made randomly and kept the same for each experiment. The models were trained for a maximum of 200 epochs with validation every 5 epochs and the model with the highest average F1 score for each split was selected. The model was implemented in Python 3.8.5 with Pytorch version 1.8.0 on a NVIDIA TITAN X Pascal (12GB) GPU with CUDA version 11.2. \

For final model assessment, we determined the classification accuracy, growth prediction sensitivity and specificity, where the metrics were averaged across all validation splits. A true positive was considered a correctly identified growing UIA, a true negative was a correctly identified stable UIA. Sensitivity and Specificity were determined using these definitions, therefore high sensitivity suggests the model is good at detecting growing UIAs and high specificity suggests the model is good at detecting stable UIAs. We plotted the mean ROC curve and calculated the mean area under the curve (AUC) for each model, as the average over all validation splits for each model.

\section{Results}
Results of the growth prediction models averaged across all validation splits are summarised in Table \ref{tab1}. Figure 2 shows ROC curves for all of the models. Roi\_model\_1, using ROI meshes and no edge mid-point co-ordinates, had the highest accuracy (0.761), F1 score (0.681) and specificity (0.883) suggesting it performs optimally for stable aneurysm detection. Uia\_model\_2, using UIA meshes and including edge mid-point co-ordinates, had the highest sensitivity for growth detection. Overall, both the second models including edge mid-point co-ordinates had higher AUC and sensitivity values but slightly lower accuracy and F1 scores. 

\begin{table}[ht]
\caption{Classification metrics for each model. F1 score is the average of F1 score for each class (growing and stable). A true positive was considered a correctly identified growing UIA, a true negative was a correctly identified stable UIA. Sensitivity and Specificity were determined using these definitions.  AUC is the area under the mean ROC curve in Figure \ref{roc_curve}. Values are provided as mean (standard deviation) across all validation splits (standard deviation)}\label{tab1}
\centering
    \begin{tabularx}{\textwidth}{|X|X|X|X|X|X|}
    \hline
    Model & Accuracy & F1 score & Sensitivity & Specificity & AUC\\
    \hline
    $uia\_model_1$  &  0.704 (0.077) &  0.617 (0.062) & 0.389 (0.048) & 0.833 (0.121) & 0.620 (0.119) \\
    $uia\_model_2$  &  0.669 (0.061) &  0.623 (0.073) & 0.573 (0.155) & 0.708 (0.075) & 0.638 (0.116) \\
    $roi\_model_1$  &  0.761 (0.017) &  0.681 (0.021) & 0.458 (0.038) & 0.883 (0.030) & 0.606 (0.056) \\
    $roi\_model_2$  &  0.713 (0.075) &  0.650 (0.077) & 0.498 (0.090) & 0.781 (0.110) & 0.622 (0.064) \\
    \hline
    \end{tabularx}
\end{table}

\begin{figure}[ht]
\includegraphics[width=\textwidth]{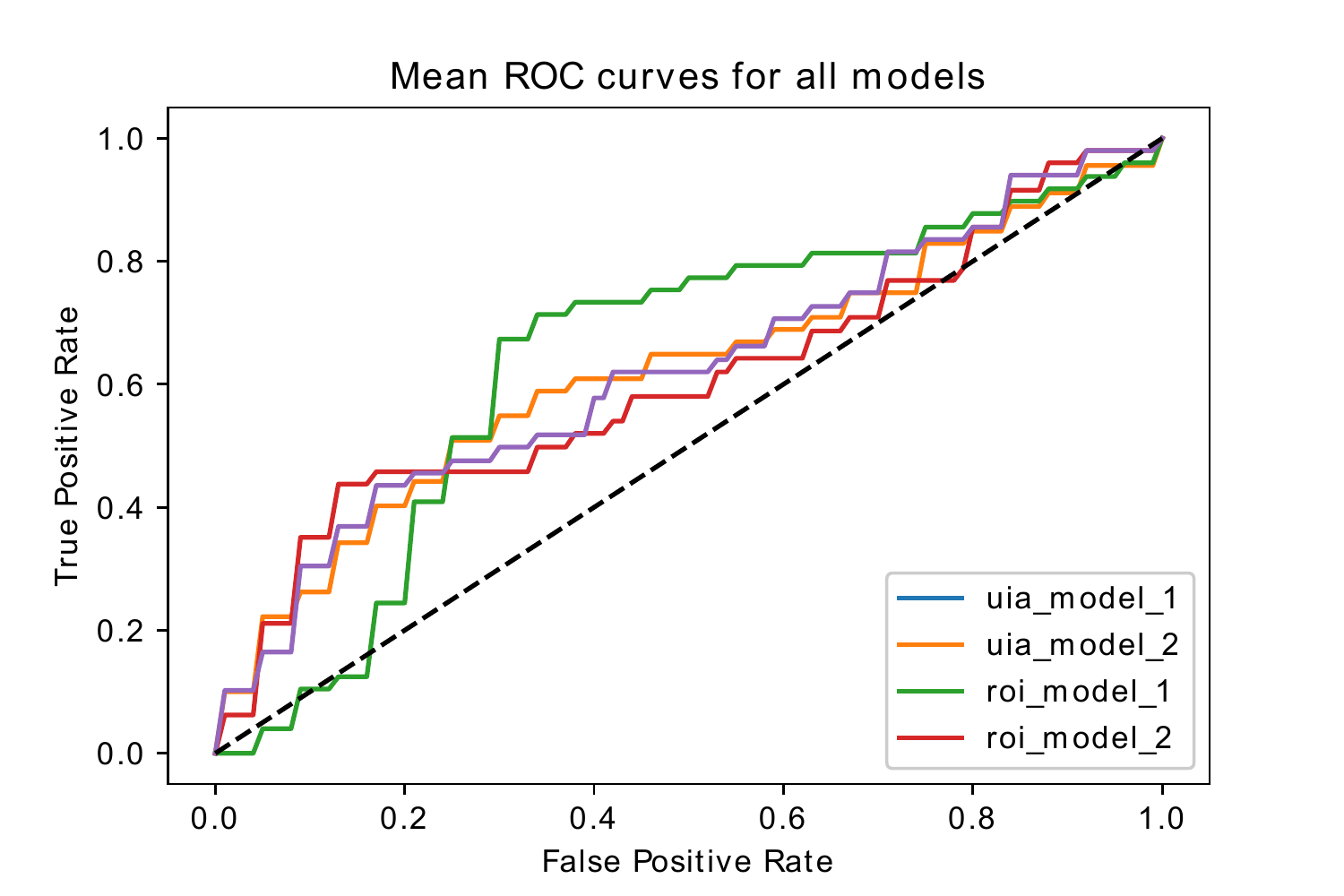}
\caption{ROC curves of all trained models for growth prediction classification. Each line is the mean of the performance across all cross validation splits for each model. The black dotted line indicates a classifier which would give random choice. } \label{roc_curve}
\end{figure}

\section{Discussion}
In this paper, we demonstrate that a future UIA growth prediction model could be developed using a mesh convolutional neural network, which considers the topology of UIAs and their parent vasculature. We found that adding edge mid-point co-ordinates as input features to the network increases the AUC and sensitivity of growth prediction but reduces the overall accuracy of the model (uia\_model\_2, roi\_model\_2). Using ROI meshes as opposed to UIA meshes alone, improved the accuracy and F1 score of the model but has a decreased AUC for growth prediction. A sensitive growth prediction model should consider using UIA meshes as input and including edge mid-point co-ordinates as input features (uia\_model\_2). \

We found using UIA meshes alone (uia\_model\_2) improved the AUC relative to using the ROI including parent vessels (roi\_model\_2). This suggests that it is the topology of the aneurysm surface itself which is high indicative of growth or stability as opposed to UIA configuration relative to parent vessels. This result is similar to previous studies, where measurements of just the UIA distinguish growing and stable UIAs \cite{Leemans2019IntracranialChanges,Leemans2019ComparingAneurysms,Timmins2022RelationshipAneurysms}. However, it is worth noting that using a ROI as opposed to the UIA mesh does not greatly reduce the performance of the method. A ROI mesh, is easier to achieve in the clinic as it requires only a click of a centre point to select the ROI. Whereas, currently the UIA meshes require accurate manual UIA segmentation. Therefore, a ROI model have more clinic applicability and has adequate performance for UIA growth prediction. \

The inclusion of input edge mid-point co-ordinate features increased the AUC and growth prediction sensitivity. This is likely because the co-ordinate provides aneurysm location information to the network. Location is a known predictor of growth \cite{Backes2017}. In the original MeshCNN paper \cite{Hanocka2019} it was commented that adding in edge co-ordinates reduced the model performance, possibly due to removing the rotation, translation and uniform scaling in-variance of the usual relative geometric input edge features. However, in real-life applications, such as in medical images, the co-ordinates give important information about the location of lesions. Therefore, the addition of these features only appears to improve the performance in this scenario. Further studies could investigate the use of relative position input features, to ensure the in-variance to rotation, translation and scaling is kept. Another possibility could be to include position/location, and potentially other known growth predictors, as global features in the final layers of the network. \

The models all had a relatively high specificity, suggesting they perform well for detecting stable UIAs. This may be useful in clinic to identify those UIAs which are stable and do not need further investigation. In our study, we had a relatively large class imbalance of only 30\% growing UIAs to 70\% stable UIAs. Although weighted loss functions and samplers were used, this does not eliminate the class imbalance. In the future, a more balanced dataset, including more growing UIAs could be used. The validation results displayed a large range in sensitivity, and it was also clear, that the model tended to over-fit relatively quickly to the training set. This is due, in part, to the heterogeneous nature of the UIAs and configurations leading to the validation sets being quite different to the training data. This could be improved by including more training and validation data. Furthermore, a larger dataset would allow for independent evaluation on a separate test set.    \

The ELAPSS growth prediction score was determined to have a c-statistic (AUC) of 0.69 in an external validation study \cite{SanchezvanKammen2019ExternalRisk}. Our model performed only slightly inferior to this (AUC = 0.64), suggesting that our model has comparable performance to current clinical prediction models. Future studies should consider combining the patient characteristics used in the ELAPSS score, with the aneurysm characteristics used in our model. 

Our proposed method did not perform as well as the method using PointNet++ put forward by Bizjak  \cite{Bizjak2021DeepGrowth} (accuracy = 82\%). This may be for a variety of reasons. Firstly, our dataset was imbalanced (30\% growing, 70\% stable) compared to the dataset they used which included more growing than stable aneurysms. Secondly, in Bizjak et al. they assess growth visually on the pre-processed 3D meshes. Instead, our model can provide a prediction for growth as is currently clinically assessed and accepted in the clinic. Future studies should investigate different definitions of growth, and as computer aided tools for UIA diagnosis and assessment continue to be developed and improve, a definition for volumetric growth should be considered \cite{Timmins2021ReliabilityAneurysms,Liu2021AStudy}. It is difficult to make a comparison of our model to the study by Liu et al. \cite{Liu2019PredictionFeatures} as they are predicting aneurysm stability, which included rupture and not just growth. Furthermore, they have a larger dataset of all aneurysms larger than 4 mm. However, future studies could investigate if our mesh based model could also predict rupture/aneurysm instability as well as growth.\

In our previous paper \cite{Timmins2022DeepDetection}, we demonstrated the mesh convolutional neural networks could be used for a modality independent UIA detection method. Based on these results, we believe that our growth prediction method could also be modality independent. This would be helpful in the clinic, where UIAs are often assessed or followed-up with different modalities such as CTA or DSA. 

\section{Conclusion}

We present a future UIA growth prediction model using a mesh convolutional neural network. We demonstrate that both UIA and ROI meshes can be used as input for such a prediction model, and that edge mid-point co-ordinates improve the growth prediction sensitivity. This model may have potential clinical use as an aid for radiologists assessing potential future UIA growth. 

\section{Acknowledgements}
We acknowledge the support from the Netherlands Cardiovascular Research Initiative: An initiative with support of the Dutch Heart Foundation, CVON2015-08 ERASE and CVON2018-02 ANEURYSM@RISK.

%
%

%
%
%
\bibliographystyle{splncs04}
\bibliography{references.bib}
%




\end{document}